\documentclass[a4,preprint,showpacs]{revtex4}
\usepackage{amsmath}
\usepackage{graphicx}
\usepackage{dcolumn}
\usepackage{bm}
\usepackage{mathrsfs}

\begin{document}
\title{
$D^-$ shallow donor near a semiconductor-metal and a semiconductor-dielectric
interface } \draft

\author{Y. L.~Hao$^{1}$, A. P. Djotyan$^{2}$, A. A. Avetisyan$^{1,2}$ and F. M.~Peeters$^{1,3}$\footnote{Electronic address: francois.peeters@ua.ac.be}}
\affiliation{$^{1}$ Department of Physics, University of
Antwerpen, Groenenborgerlaan 171, B-2020 Antwerpen, Belgium }
\affiliation{$^{2}$ Yerevan State University, Department of
Physics, A. Manoogian 1, Yerevan 0025, Armenia,}
\affiliation{$^{3}$ Departamento de F\'isica, Universidade Federal
do Cear\'a, Caixa Postal 6030, Campus do Pici, Fortaleza 60455-900, CE, Brazil}

\date{\today}

\begin{abstract}
The ground state energy and the extend of the wavefunction of a negatively charged donor ($D^-$)
located near a semiconductor-metal or a semiconductor-dielectric interface is obtained. We apply
the effective mass approximation and use a variational two-electron wavefunction that takes into
account the influence of all image charges that arise due to the presence of the interface, as
well as the correlation between the two electrons bound to the donor. For a semiconductor-metal
interface, the $D^-$ binding energy is enhanced for donor positions $d>1.5a_B$ ($a_B$ is the
effective Bohr radius) due to the additional attraction of the electrons with their images. When
the donor approaches the interface (i.e. $d<1.5a_B$) the $D^-$ binding energy drops and eventually
it becomes unbound. For a semiconductor-dielectric (or a semiconductor-vacuum) interface the $D^-$
binding energy is reduced for any donor position as compared to the bulk case and the system
becomes rapidly unbound when the donor approaches the interface.

\end{abstract}


\maketitle

\section{Introduction}
The negatively charged donor center, also called $D^-$ system (a hydrogenic donor with a second
electron bound to it) has received a lot of interest in the past both from researchers in
astrophysics (where it is known as the H$^-$ ion~\cite{Chand}) and semiconductor
physics~\cite{Hill}. Furthermore, the $D^-$ center is one of the simplest ``many-body'' electronic
system that can also be used as a model system to test how well certain theories are able to
include electron-electron (e-e) correlations. It has been shown that in the absence of electric
and magnetic fields the $D^-$ has only one bound state~\cite{Hill}. Experimentally, $D^-$ states
have been observed in bulk semiconductors~\cite{ExpBulkSemi}, in quantum wells~\cite{ExpQW}, and
in superlattices~\cite{Huant}. Tunneling through the $D^-$ state was reported in
Ref.~\onlinecite{Lok} for a $D^-$ confined in a double-barrier resonant tunneling device. The
$D^-$ resonance appears in addition to the known resonance due to tunneling through the ground
state of the neutral donor $D^0$. It was found that in high magnetic fields the amplitude of the
$D^-$ resonant peak becomes significantly larger as compared to the amplitude of the $D^0$
peak~\cite{Lok}.

Theoretically, the $D^-$ center has been investigated using the variational method within the
effective mass approach: i) for the bulk case~\cite{Larsen1}, ii) the two-dimensional
case~\cite{Larsen2,Larsen3}, and iii) for the superlattice case~\cite{Shiprb}. In a
double-quantum-well device a neutral donor $D^0$ at the center of the quantum well in the presence
of magnetic field can bind a second electron in the other well (a spatially separated $D^-$
center) as shown in Ref.~\cite{IKM}. In the case of the remote $D^-$ center, where donor and
electrons are located in different quantum wells, strong electron-electron correlations can give
rise to magnetic-field-induced angular-momentum transitions~\cite{IKM,Riva}. It has been predicted
that electron-phonon interaction in a weakly polar semiconductor leads to a substantial increase
of the $D^-$ binding energy~\cite{Shiprb2}. The negatively charged donor center was used as an
approximate model system for a trion (charged exciton) to explain the experimental behavior of the
two-dimensional electron gas in a quantum well in high magnetic field at high-laser
power~\cite{Hayne}.

In the last years, there has been renewed interest in the study of dopants in semiconductors due
to the possibility to dope the material in a controlled way and to tailor the electronic
properties in order to create new functional devices. Because of the increased miniaturization,
the dopant atoms appear closer and closer to interfaces~\cite{Koenraad}. Recently, using STM, the
binding energy of individual dopants close to a semiconductor interface was measured and found to
be substantially increased~\cite{Koenraad}. In a recent transport experiment on a nanowire
surrounded by a metallic gate it was suggested~\cite{Sellier} that signatures of the $D^-$ state
were observed. Due to the closeness of the metallic gate to the donor it was argued that the
metallic gate screens the repulsive e-e interaction which should lead to a larger binding energy.
Here we will show that this expectation is only correct if the donor is not too close to the
interface. For very close proximity of the donor to the interface the $D^-$ binding energy drops
and can even become negative, i.e. unbound $D^-$ system. A second motivation for the present study
is that the $D^-$ system can serve as an entangled pair of electrons which currently is of great
interest for quantum information applications~\cite{Bao}. The quantum control of a $D^0$ near a
semiconductor-dielectric interface and its possible application for quantum computing has been
discussed in Ref~\cite{Calderon}.

In this paper we study the spin-singlet state of a $D^-$ system near a semiconductor-metal (and
semiconductor-dielectric) interface within the effective mass approach. As compared to the 3D
situation the present problem differs in the following two aspects: 1) the many-particle
wavefunction is zero at the interface, and 2) due to the dielectric mismatch at the interface
image charges are induced that results in a complicated multi-center Coulomb problem. For the
ground state trial function we used a Chandrasekhar type space symmetric wave function which we
modified in order to satisfy the boundary condition on the semiconductor-insulator interface, and
to take into account the contribution of the Coulomb interaction with the image charges in the
system. To describe the interaction of the electrons with the images in the insulator or in the
metallic gate, we add additional terms to the Chandrasekhar trial function~\cite{Chand}. For the
case of a single electron bound to a donor near a metallic or dielectric interface, i.e. the
$D^{0}$ problem, similar terms were introduced recently in Ref.~\onlinecite{Hao} which we modified
slightly in order to obtain even better results for $d>a_B$. In the present paper, we have
calculated the ground state energy of the $D^-$ system as a function of the donor position with
respect to a semiconductor-metal, a semiconductor-dielectric and a semiconductor-vacuum interface.

The present paper is organized as follows. In Sec.~\ref{sec2} we present the two electron
Hamiltonian and propose trial wave functions for the $D^-$ center. In Sec.~\ref{sec3} the ground
state energy of the $D^-$ center near different interfaces is studied as a function of the
position of the donor. The extend of the wavefunction and its average position in the direction
parallel and perpendicular to the interface were also calculated. Our conclusions and a summary of
our results are presented in Sec.~\ref{sec4}.

\section{The formalism and the variational wave function} \label{sec2}
The Hamiltonian of the $D^-$ system, consisting of a donor at a position ${\bf{r}}_d  = (0,0,d)$
near a semiconductor-metal (semiconductor-dielectric) interface and two electrons is, in
cylindrical coordinates, given by the expression
\begin{equation}
H = H_1  + H_2  + U(\vec{\bf{r}}_1,\vec{\bf{r}}_2),
\label{eq_H}\end{equation} %
where
\begin{equation}
H_i  =  - \frac{1}{2}\left[ {\frac{{\partial ^2 }}{{\partial \rho
_i^2 }} + \frac{1}{{\rho _i }}\frac{{\partial }}{{\partial \rho
_i}} + \frac{1}{{\rho _i^2 }}\frac{{\partial ^2 }}{{\partial
\theta _i^2 }} + \sigma \frac{{\partial ^2 }}{{\partial z_i^2 }}}
\right]+U^0(\vec{\bf{r}}_i) \label{eq_H_i}\end{equation} is the Hamiltonian
of a neutral $D^0$ center near an interface, with
\begin{equation}
U^0(\vec{\bf{r}}_i)=\frac{Q}{{4z_i }} - \frac{1}{{\sqrt {\rho _i ^2  + (z_i - d)^2 } }} -
\frac{Q}{{\sqrt {\rho _i ^2  + (z_i  + d)^2 } }}, \label{eq_U_i}\end{equation} the Coulomb
interaction terms and where dimensionless units of the effective Bohr radius $ a_B  = {{\hbar ^2
\varepsilon _1 } \mathord{\left/
 {\vphantom {{\hbar ^2 \varepsilon _1 } {m_ \bot  }}} \right.
 \kern-\nulldelimiterspace} {m_ \bot  }}e^2$ and twice the effective Rydberg energy $
2 R^ *   = {{\hbar ^2 } \mathord{\left/
 {\vphantom {{\hbar ^2 } {m_ \bot  }}} \right.
 \kern-\nulldelimiterspace} {m_ \bot  }}a_B ^2$ were used and
$\sigma=m_\bot/m_{||}$
is the ratio between the transverse and longitudinal effective mass (in our
numerical calculations we assume $\sigma=1$).

The first term in $U^0(\vec{\bf{r}}_i)$ describes the interaction between an electron and its
image, the second arises due to the attractive interaction between an electron and the donor, and
the third term is due to the interaction between an electron and the donor image (as well as the
donor and the electron image). See Fig.~\ref{fig_interaction} for a schematic representation of
the different Coulomb terms in case of a semiconductor-metal interface for which $Q=-1$ (i.e.
$\varepsilon _{O}\rightarrow \infty$). In Eq.~(\ref{eq_U_i}) the image charge is given by $Q =
(\varepsilon _{S} - \varepsilon _{O} )/(\varepsilon _{S} + \varepsilon _{O} )$ with
$\varepsilon_{S}$ ($\varepsilon_{O}$) the permittivity of the semiconductor (dielectric). For the
case of a semiconductor-metal interface we assume a very thin oxide layer between the
semiconductor and the metal and its only effect is to prevent the electron to penetrate into the
metal, i.e. it provides a very high potential barrier. The two-electron Coulomb potential has the
following form
\begin{equation}
U(\vec{\bf{r}}_1,\vec{\bf{r}}_2)=U_{ee}(\vec{\bf{r}}_1,\vec{\bf{r}}_2)+U_{ei}(\vec{\bf{r}}_1,\vec{\bf{r}}_2), \label{eq_U}
\end{equation}
with
\stepcounter{equation}
\begin{align}
&U_{ee}(\vec{\bf{r}}_1,\vec{\bf{r}}_2)=\frac{1}{{\sqrt {\rho _1 ^2  + \rho _2 ^2  - 2\rho _1 \rho _2\cos (\theta _1  - \theta _2 ) + (z_1  - z_2 )^2 } }}, \tag{\theequation a} \label{eq_Uee}\\
&U_{ei}(\vec{\bf{r}}_1,\vec{\bf{r}}_2)=\frac{Q}{{\sqrt{\rho _1 ^2  + \rho _2 ^2  - 2\rho _1 \rho _2 \cos (\theta _1  - \theta _2 ) + (z_1  + z_2 )^2 } }}, \tag{\theequation b} \label{eq_Uei}
\end{align}
where $U_{ee}$ describes the electron-electron interaction and $U_{ei}$ is the interaction between
an electron and the image of the other electron.

\begin{figure}[tb]
\begin{center}
\includegraphics*[width=\columnwidth]{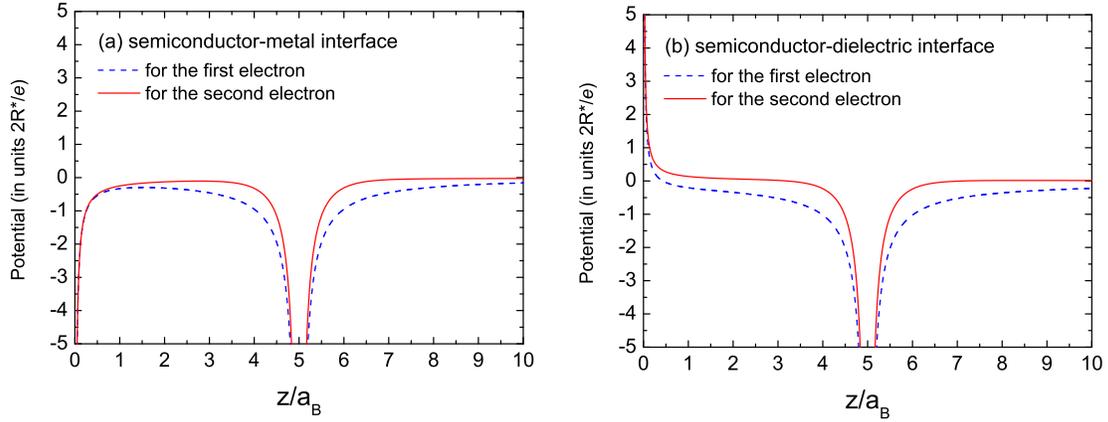}
\end{center}
\caption{(Color online) The mean-field potential energy along the z-axis (with $\rho=0$) for a donor located at
$d/a_B=5$ in case of (a) a semiconductor-metal and (b) a semiconductor-dielectric
($\varepsilon_s=11.9$, $\varepsilon_d=3.4$) interface. We show the result for the potential of a
single electron (dashed blue curve) and for the electrostatic potential due to the unperturbed
$D^0$ system with the electron in its ground state (solid red curve).}
\label{fig_WholeD0Potential}\end{figure}

The potential energy for the $D^0$ electron along the $z$-direction for $\rho=0$ is shown by the
dashed curve in Fig.~\ref{fig_WholeD0Potential}(a) for the semiconductor-metal interface and in
Fig.~\ref{fig_WholeD0Potential}(b) for the semiconductor-dielectric interface. The electron is
subject to a $-e/|\vec{\bf{r}}-\vec{\bf{r_d}}|$ Coulomb potential near the donor and the potential
due to the image charges. When an electron is bound to the positive donor the system becomes
neutral and this electron will screen the donor Coulomb potential for a second electron. The
mean-field potential seen by the second electron is given by
\begin{equation}
U^0(\vec{\bf{r}}_2)+\langle\psi_{D^0}(\vec{\bf{r}}_1)|U(\vec{\bf{r}}_1,\vec{\bf{r}}_2)|
\psi_{D^0}(\vec{\bf{r}}_1)\rangle \label{eq_UD0}
\end{equation}
where $\psi_{D^0}(\vec{\bf{r}})$ is the normalized ground-state wavefunction of the $D^0$ electron
which we obtained previously in Ref.~\onlinecite{Hao}. This potential is plotted in
Figs.~\ref{fig_WholeD0Potential}(a,b) by the solid curve and it is immediately clear that the
second electron will feel a strongly screened donor potential. The $1/r$ Coulomb potential is
replaced~\cite{Larsen2} by the screened potential $(1+r)e^{-2r}/r$. As a result the second
electron can at most only be very weakly bound. Solving this mean-field problem for the second
electron will strongly underestimate the $D^-$ binding energy because it neglects: i) exchange,
and ii) electron-electron correlation. The latter will lead to a polarization of the neutral $D^0$
system.

In order to account for exchange and correlation we introduce a Chandrasekhar type space symmetric
wave function (because we consider the spin single state, i.e.
$\psi(\vec{r}_1,\vec{r}_2)=\psi(\vec{r}_2,\vec{r}_1)$) which also takes into account the
interaction with the image charges in the system, and satisfies the boundary condition at the
interface:
\begin{equation}
\psi(\vec{\bf{r}}_1,\vec{\bf{r}}_2) = N\,p(z_1)\,p(z_2)\,[g (\vec{\bf{r}}_1,\vec{\bf{r}}_2) +
g(\vec{\bf{r}}_2,\vec{\bf{r}}_1)] \varphi(\vec{\bf{r}}_1,\vec{\bf{r}}_2),
\label{eq_psi}\end{equation}
where %
\stepcounter{equation} %
\begin{align}
p(z_i)&=z_i/(1+\alpha z_i),\tag{\theequation a} \\ %
g(\vec{\bf{r}}_1,\vec{\bf{r}}_2)&=\exp ( - \lambda _1 r_1 - \lambda _2 r_2 ) \exp (- \beta _1
p(z_1)  - \beta _2 p(z_2)), \tag{\theequation b} \label{eq_g} \\ %
\varphi(\vec{\bf{r}}_1,\vec{\bf{r}}_2)&=1+\delta r_{12}, \tag{\theequation c} \label{eq_varphi} %
\end{align}
with $r_i = \sqrt {\rho _i^2  + (z_i  - d)^2 }$ the distance between the electron and the donor,
$r_{12}=\sqrt {\rho _1 ^2  + \rho _2 ^2  - 2\rho _1 \rho _2\cos (\theta _1  - \theta _2 ) + (z_1 -
z_2 )^2}$ is the distance between the two electrons and $N$ is a normalization constant. In our
previous work~\cite{Hao} on the neutral donor problem we found that introducing an asymmetry in
$r_i$, i.e. $r_i=\sqrt{\rho^2_i+\gamma (z_i-d)^2}$, with $\gamma$ an extra variational parameter,
improved the binding energy only with 2.7\% for $d/a_B=1$ and with less than 1\% for $d/a_B=5$.
Therefore, in order to limit the number of variational parameters we took $\gamma=1$. The factor
$p(z_i)$ in front of the wave function~(\ref{eq_psi}) guarantees that the boundary condition $\psi
(z_i  = 0) = 0$ is satisfied for each electron and that for large $d$-values the 3D result can be
recovered (because then $p(z_i)$ approach the constant value $1/\alpha$). The wave function
contains 6 variational parameters: $\lambda_1$, $\lambda_2$, $\beta_1$, $\beta_2$, $\delta$ and
$\alpha$. The symmetric combination of hydrogen-like factors $\exp(-\lambda_i r_i)$ in
Eq.~(\ref{eq_g}), takes into account the interactions of the two electrons with the donor, as was
proposed in Ref.~\onlinecite{Chand}, and $\varphi(\vec{\bf{r}}_1,\vec{\bf{r}}_2)$ describes the
electron-electron correlation. We introduced the combination of exponential factors $f(z_i) = \exp
( -\beta_i\,p(z_i) )$ that describes the overall interaction of each electron with its image, as
well as with the images of the donor and the other electron. Similar functions for the
electron-electron repulsive interaction (but with $\alpha=0$) were used previously for the $D^-$
system in quantum wells~\cite{Shiprb} and for the two-electron parabolic quantum
dot~\cite{Avetisjpm}.

For a neutral electron-donor $D^{0}$ problem near an
interface~\cite{Hao} we took $\alpha=0$ and thus $p(z)=z$ which
resulted in very good agreement between the results of the
variational method and a numerical 'exact' finite-element
solution. Although the energy was found to be accurate it is clear
that the bulk wavefunction is not recovered for
$d\rightarrow\infty$, which should be spherical symmetric around
the donor. This is the reason why we introduced
$p(z_i)=z_i/(1+\alpha z_i)$ which for $z\rightarrow\infty$
approaches the constant value $1/\alpha$ and makes the
wavefunction spherical symmetric. In Table~\ref{table_fz} we
compare the obtained variational energy for the $D^-$ using
$p(z_i)=z_i$ and $p(z_i)=z_i/(1+\alpha z_i)$; we see that the
latter gives lower energy for all $d$-values. Thus this extra
variational parameter has only a small influence on the energy (we
checked that the same conclusion holds for the $D^0$ problem). But
we found that quantities as, e.g., the average electron position
behave much better at large $d$ if we include $\alpha$ as a
variational parameter. For $d\rightarrow\infty$ we have
$p\rightarrow 1/\alpha$ and our variational wavefunction reduces
to the one proposed by Chandrasekar~\cite{Chand} which has been
shown to result in accurate values for the bulk $D^-$ binding
energy.

\begin{table}[tb]
\begin{tabular}{cccc}
\hline %
\hline %
                              & \multicolumn{3}{c}{$E/2R^*$} \\
\raisebox{2.3ex}[0pt]{$d/a_B$}& $\qquad \quad p(z_i)=z_i \qquad \quad $& $p(z_i)=z_i/(1+\alpha z_i)$ & $\qquad \text{asymmetric e-e}$ \qquad \\
\hline %
2                             & -0.40883                               & -0.41464                                & -0.41464 \\
3                             & -0.49106                               & -0.49783                                & -0.49783 \\
4                             & -0.51680                               & -0.52180                                & -0.52186 \\
5                             & -0.52518                               & -0.52885                                & -0.52891 \\
6                             & -0.52768                               & -0.53061                                & -0.53067 \\
7                             & -0.52811                               & -0.53059                                & -0.53063 \\
8                             & -0.52787                               & -0.52998                                & -0.53001 \\
9                             & -0.52747                               & -0.52922                                & -0.52924 \\
10                            & -0.52709                               & -0.52850                                & -0.52852 \\
\hline %
\hline %
\end{tabular}
\caption{The ground state energy of a $D^-$ center near a semiconductor-metal interface at
different donor-interface distance $d$, calculated by using two different $p(z_i)$ terms in the
wave function, Eq.~(\ref{eq_psi}). In the last column we present the results for $\alpha \neq 0$
and including the asymmetric e-e correlation term, Eq.~(\ref{eq_deltaZ}), in the variational wave
function.} \label{table_fz}
\end{table}

Due to the presence of the interface the e-e correlation term is not necessarily circular
symmetric. In order to check the influence of this asymmetry on the energy we introduced an
additional variational parameter $\delta_z$ in the e-e correlation part of Eq.~(\ref{eq_psi}),
\begin{equation}
\varphi_D(\vec{\bf{r}}_1,\vec{\bf{r}}_2)=1+\delta \sqrt {\rho _1 ^2  + \rho _2 ^2  - 2\rho _1 \rho
_2\cos (\theta _1  - \theta _2 ) + \delta_z (z_1  - z_2 )^2}. \label{eq_deltaZ}
\end{equation}
The ground state energy of the $D^-$ near a semiconductor-metal
interface calculated by using $\varphi_D$ in the wavefunction
Eq.~(\ref{eq_psi}) is presented in the last column of
Table~\ref{table_fz}. Notice that allowing the e-e correlation to
be anisotropic does not have a significant influence on the
energy, the differences between the energy calculated with or
without $\delta_z$ are very small (less than 0.02\%). Therefore,
in the following we did not include $\delta_z$ in our
calculations.

The $D^-$ binding energy is defined as follows:
\begin{equation}
E_b(d)=E_{D^{0}}-E_{D^{-}}, \label{bind}
\end{equation}
where $E_{D^{0}}$ is the ground state energy of the $D^0$ system~\cite{Hao}. Eq.~(\ref{bind})
gives the energy that is needed to remove one of the electrons from the donor to infinity. This
definition corresponds to the one used for a 3D $D^-$ system. For a metallic interface, another
definition of the binding energy can be used:
\begin{equation}
E_b(d)=E_f+E_{D^{0}}-E_{D^{-}}, \label{bind_mtl}
\end{equation}
where $E_{f}/2R^*=-0.0312$ is the binding energy of an electron with its image near a metallic
interface~\cite{Hao,Loudon}. This is the energy needed to remove one electron from the $D^-$ and
bring it to the interface.

The expectation values for the position of electron $i$ in the $z$-direction and in the $\rho$
plane are calculated as $\langle\rho_i\rangle=\int {\rho_i \, \chi_{12}^2 } \, d{\bf{r}}$ and
$\langle z_i-d \rangle = \int {(z_i-d) \,\chi_{12} ^2 } \, d{\bf{r}}$, respectively, using only
the first part in the total trial function, i.e. $\chi_{12}=N\,p(z_1)p(z_2)\,
g(\vec{\bf{r}}_1,\vec{\bf{r}}_2) \varphi(\vec{\bf{r}}_1,\vec{\bf{r}}_2)$. This allows us to
discriminate between the two electrons, one is closely bound to the positive impurity while the
other one is very weakly bound. Notice that thanks to our variational wavefunction we are able to
separate 'artificially' the two electrons. But the variational wavefunction itself considers the
two electrons to be indistinguishable because
$\psi(\vec{\bf{r}}_1,\vec{\bf{r}}_2)=\psi(\vec{\bf{r}}_2,\vec{\bf{r}}_1)$.

\section{The ground state energy of the $D^-$ center} \label{sec3}
\noindent \textbf{A. Semiconductor-metal interface}

\begin{figure}[tb]
\begin{center}
\includegraphics*[width=12cm]{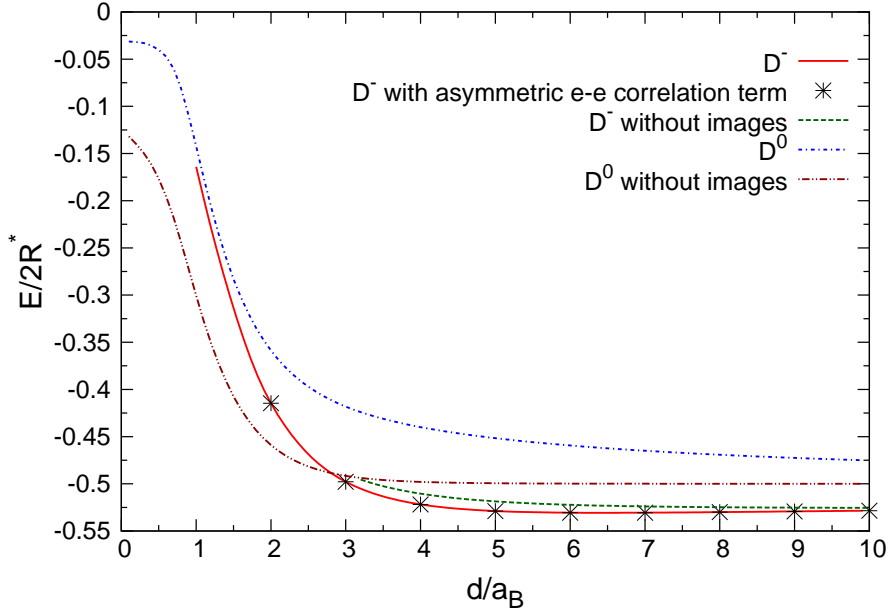}
\end{center}
\caption{(Color online) The ground state energy (solid red curve)
for a $D^-$ system near a semiconductor-metal interface vs. the
distance of the donor $d/a_B$ from the interface. Dashed green
curve is the results when the image charges are neglected and by
black stars we indicate the results when an additional variational
parameter $\delta_z$ is included in the e-e correlation term (see
Eq.~(\ref{eq_deltaZ})). For comparison the energies for a neutral
donor ($D^0$) (dot-dashed blue curve) and in the absence of the
image charges (dot-dot-dashed dark red curve) are also shown. }
\label{gr_en_m}\end{figure}

The ground state energy, in units of $2R^*$, is shown in Fig.~\ref{gr_en_m} as a function of the
distance of the donor from the interface $d/a_B$. The $D^-$ energy exhibits a shallow minimum
around $d\sim6.5a_B$ and approaches the bulk result $E_{bulk}/2R^*=-0.5259$ from below for $d
\rightarrow \infty$. For $d<6.5a_B$ the $D^-$ energy is a decreasing function of $d$ which is
mostly a consequence of the extra constraint that the wavefunction has to be zero at the
interface. This can be inferred from the result without image charges (dashed green curve in
Fig.~\ref{gr_en_m}). The outer electron of the $D^-$ center becomes bound for $d\geq 1.1a_B$. We
considered the second electron bound if the variational parameter $\lambda_2$ is larger than zero
and the average position of the electron is not too far from the donor, i.e. $\langle z_2-d
\rangle \ll 10a_B$. For a semiconductor-metal interface, we have two positive image charges from
the electrons and only one negative image charge from the donor, so a strengthened electron bound
state is expected. This is confirmed by the fact that if we ignore the image charges the second
electron becomes only bound when $d \geq 3.2 a_B$ (see the dashed green curve in
Fig.~\ref{gr_en_m}).

\begin{figure}[tb]
\begin{center}
\includegraphics*[width=12cm]{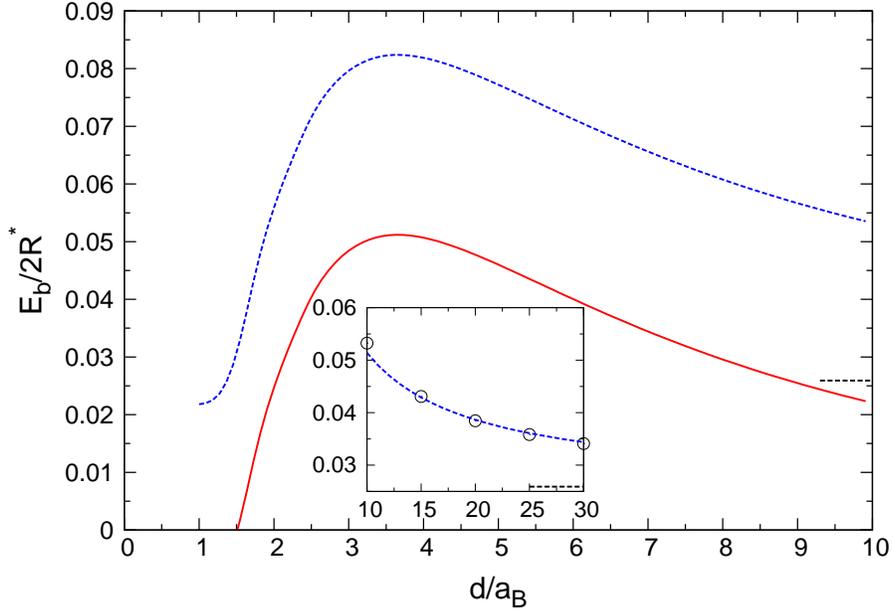}
\end{center}
\caption{(Color online) The binding energy (solid red curve) for a $D^-$ center near a
semiconductor-metal interface vs. the donor-interface distance $d/a_B$. The binding energy in the
absence of $E_f$ (the energy of an electron bound to the interface) is shown by the dashed blue
curve. The horizontal dashed black line shows the bulk limit of the binding energy. The inset
shows the behavior of the binding energy for large $d$ (black circles) which is fitted to the
curve $E_b/2R^*=0.0259+0.255a_B/d$ (dashed blue curve).} \label{bind_met}\end{figure}

The binding energy is plotted in Fig.~\ref{bind_met} using the two different definitions of the
binding energy (Eqs.~(\ref{bind}) and (\ref{bind_mtl})). A clear maximum is found for $d \approx
3.8 a_B $ beyond which the binding energy slowly decreases to its $d \rightarrow \infty$ value.
This algebraic decrease is due to the interaction with the image charges and is largely a
consequence of the $d$-dependence of the $D^0$ energy. For the definition of the binding energy
with $E_f=0$ (blue dashed curve in Fig.~\ref{bind_met}) the bulk $D^-$ binding energy
$E_{b,bulk}/2R^*=0.0259$ is reached from above. Notice that for $d=3.8a_B$, $E_{b}/2R^*=0.0823$
which is a factor 3.18 larger than the bulk $D^-$ binding energy. The large $d$-range is shown in
the inset and we found that it can be fitted by the curve $E_b/2R^*=0.0259+0.255a_B/d$.

\begin{figure}[tb]
\begin{center}
\includegraphics*[width=12cm]{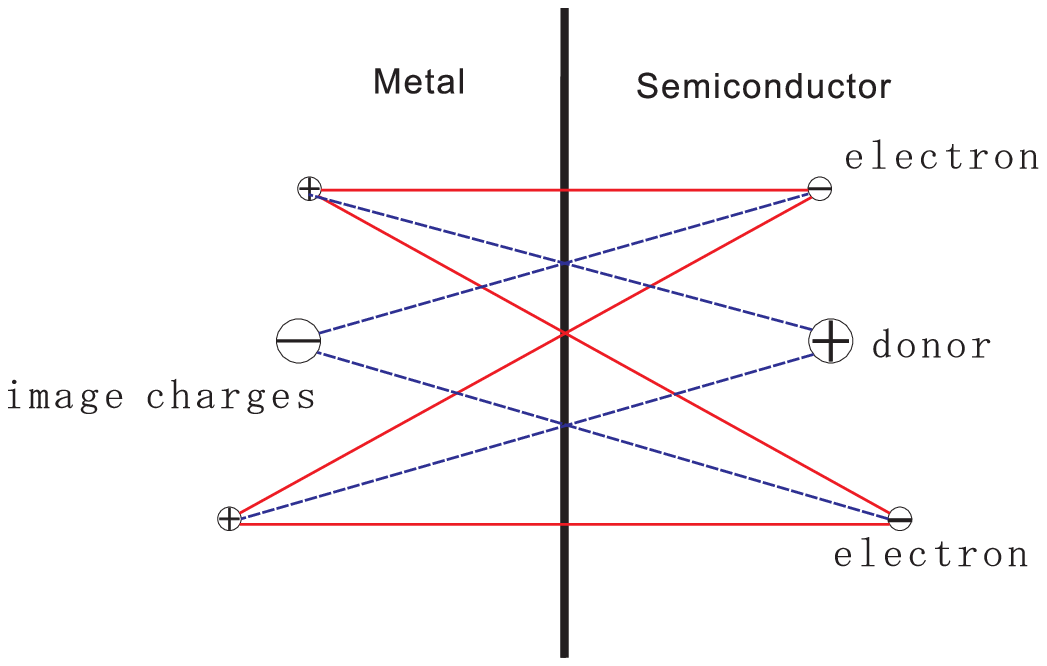}
\end{center}
\caption{(Color online) Schematic picture of the interaction between the $D^-$ system and its
image charges for the case of a semiconductor-metal interface. Dashed (solid) lines indicate the
repulsive (attractive) interactions between the electrons, the donor and their images.}
\label{fig_interaction}\end{figure}

It is remarkable that in the $D^-$ case the effect of the image charges on its energy almost
cancels out for intermediate $d$, while this is not so for the neutral $D^0$ system. In our
previous work~\cite{Hao} we found that in the case of a semiconductor-metal interface the
contribution of the image terms to the energy of a neutral $D^0$ center is given by $\Delta E =
1/4d$ for large $d$. This is the sum of the contributions arising due to the interaction of the
electron with its image (about $-1/4d$) and the interactions of the electron with the donor image
as well as the electron image with the donor (each one is about $1/4d$). In
Fig.~\ref{fig_interaction} we show by dashed lines the repulsive interactions between the
electrons and the donor image as well as between the donor and the images of the electrons (each
line is characterized by the Coulomb energy $1/4d$) and by solid lines the attractive interactions
between the electrons and their images (each with the energy $-1/4d$). Now, in the case of the
$D^-$, due to the presence of the two electrons (see Fig.~\ref{fig_interaction}), we have twice
the energy shift of the single electron problem $2\Delta E = (1/4d)$, and there is also twice a
negative shift due to the interaction of each electron with the image of the other electron
$2(-1/4)d$ (see Eq.~(\ref{eq_Uei})). As a result these energy shifts compensate each other, and we
find that the energy for $d > 10a_B$ is equal to the energy of the $D^-$ without image charges and
is very close to the energy $E_{bulk}/2R^*=-0.5259$ of $D^-$ in the bulk~\cite{Chand}. The same
compensation takes place in the case of a semiconductor-dielectric interface where the large $d$
contribution of image terms in the energy of the neutral D$^0$ center is about $-Q/4d$

\begin{figure}[tb]
\begin{center}
\includegraphics*[width=10cm]{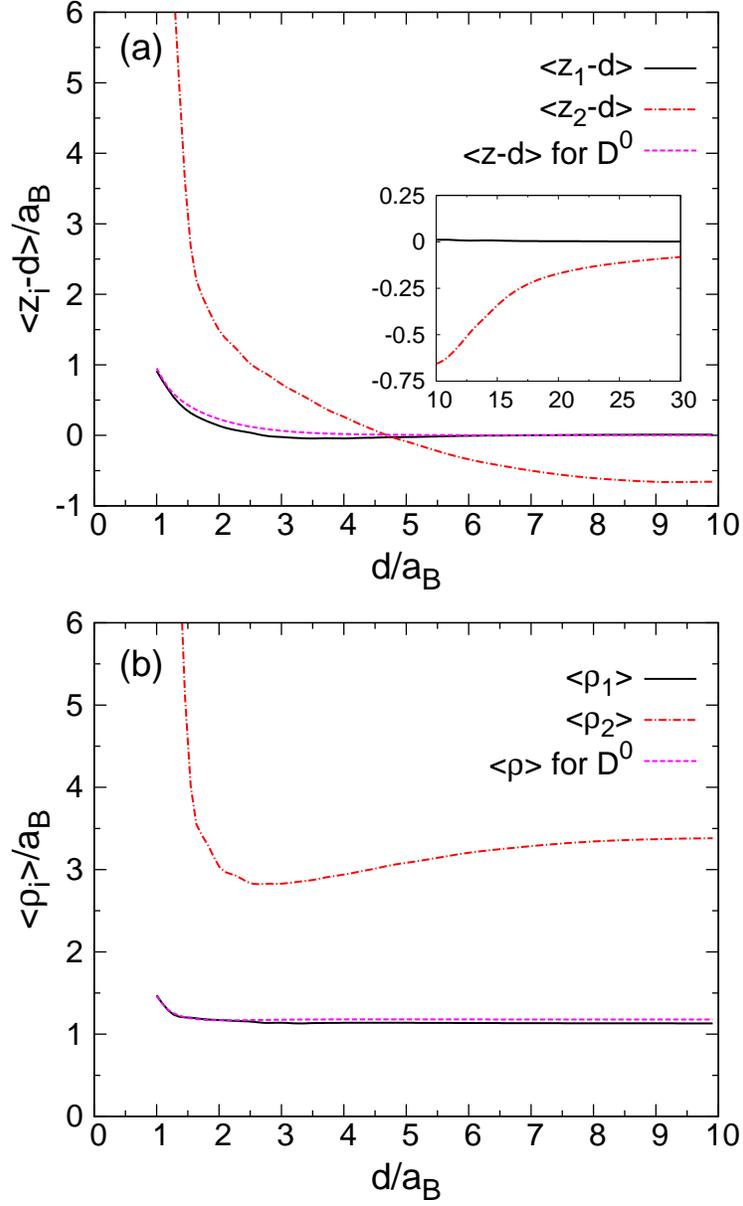}
\end{center}
\caption{(Color online) For the semiconductor-metal interface: (a) The average distance between
the electrons and the donor along the $z$ direction vs. the donor position $d$. (b) The average
value of the electron positions along the $\rho$ direction vs. the donor position $d$. The inset
of figure (a) shows the behavior of $\langle z_i-d \rangle$ at large $d$.}
\label{fig_avpos_mtl}\end{figure}

\begin{figure}[tb]
\begin{center}
\includegraphics*[width=13cm]{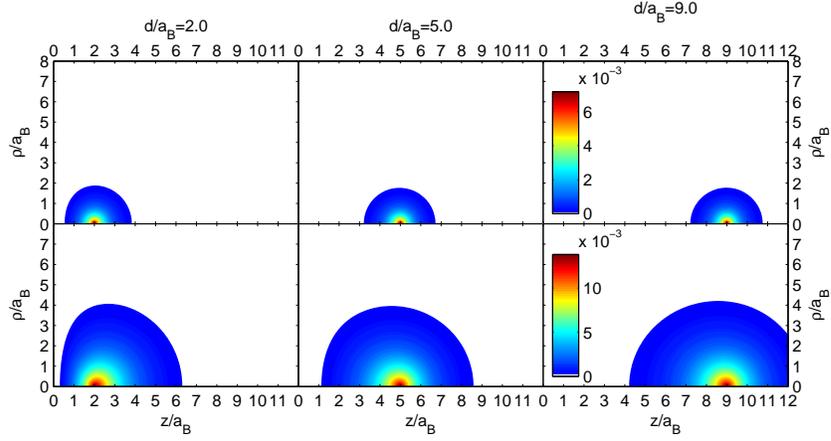}
\end{center}
\caption{(Color online) The electron probability density of the $D^-$ center near a
semiconductor-metal interface for different donor distances $d/a_B=2$, 5 and 9 when the other
electron is fixed to its average position (see the corresponding average values in
Fig.~\ref{fig_avpos_mtl}) as well as the angle between two electrons $(\theta_1-\theta_2)$ is set
to its average value. The upper three figures correspond to the inner electron density (when the
outer electron is fixed) , whereas the lower three figures are densities of the outer electron.
Blue (red) area represents low (high) probabilities and the white area represents almost zero
probability.}
\label{fig_eDen_mtl}\end{figure}

\begin{figure}[tb]
\begin{center}
\includegraphics*[width=12cm]{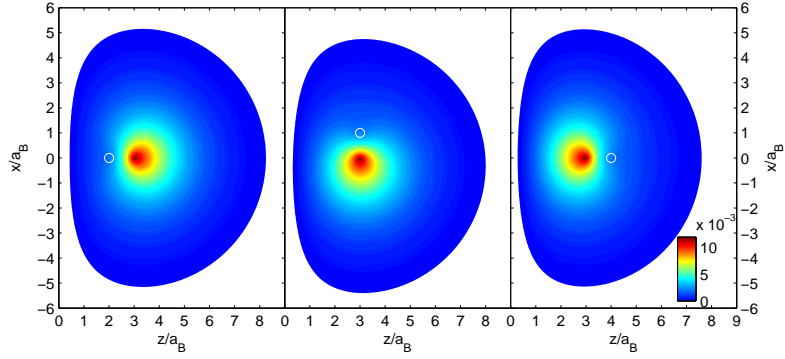}
\end{center}
\caption{(Color online) The conditional electron probability density of the outer electron where
the inner electron is fixed when the $D^-$ center near a semiconductor-metal interface with
$d/a_B=3$. The white circles marked out the location of the inner electron, that is (from left to
right, in the form ($x/a_B$, $y/a_B$, $z/a_B$)) (0, 0, 2), (1, 0, 3) and (0, 0, 4). }
\label{fig_eDen_mtl_fixed_r0}
\end{figure}

The average position of the electrons in the $z$-direction and their extend in the plane parallel
to the interface are shown in Figs.~\ref{fig_avpos_mtl}(a) and \ref{fig_avpos_mtl}(b),
respectively. Notice that one of the electrons, also called the inner electron, follows very
closely the behavior of the electron bound in the neutral $D^0$ system. This is very similar to
what was found previously for the case of a bulk $D^-$~\cite{Larsen1}. The second electron, called
the outer electron, is more extended in the $\rho$-plane, i.e. its average value $\langle \rho_2
\rangle$ is about three times larger than for the inner electron. Notice that $\langle \rho_2
\rangle$ for the outer electron increases rapidly with decreasing $d$ when $d<2a_B$ signaling a
rapid decrease of the binding energy and ultimately an unbinding of the outer electron. For
$d>4.75a_B$ the outer electron is attracted towards the interface, i.e. $\langle z_2 \rangle<d$,
as a consequence of the image charges which is responsible for the enhanced $D^-$ binding energy.
For smaller $d$ values we have $\langle z_2 \rangle>d$ and the outer electron is pushed away from
the interface mostly as a consequence of the boundary condition at the interface. This behavior is
also illustrated in Fig.~\ref{fig_eDen_mtl} where we show the contour plots of the electron
density of the outer and inner electron, i.e. $|\psi_2(z_2,\rho_2)|^2=p(z_1)^2 p(z_2)^2
g(\vec{\bf{r}}_1,\vec{\bf{r}}_2)^2\varphi(\vec{\bf{r}}_1,\vec{\bf{r}}_2)^2$ with $r_1$, $\rho_1$
and $(\theta_1-\theta_2)$ taken as their average value, for $d/a_B=2$, 5, 9. Notice that the
electron distribution is asymmetric when the donor is close to the interface and that in such a
case a large part of the distribution is found with $z>d$. In order to illustrate the effect of
the electron-electron correlation we show in Fig.~\ref{fig_eDen_mtl_fixed_r0} the conditional
probability $P(\vec{r},\vec{r}_0)=\left|\psi(\vec{r},\vec{r}_0)\right|^2$. This is the probability
to find an electron at position $\vec{r}$ when the other electron is fixed at position
$\vec{r}_0$. We fix one of the electrons close to the donor (i.e. it is the inner electron) and
put it in three different positions with respect to the interface-donor axis. Notice that the
electron: 1) has the highest probability to be close to the donor, 2) it is repelled by the fixed
electron, and 3) it has a non-zero probability to be located at $\vec{r}_0$. The reason is that
the two electrons have opposite spin and therefore the Pauli exclusion principle is not
applicable.
\newline

\noindent \textbf{B. Semiconductor-dielectric interface}

In this subsection we investigate the ground state energy of a $D^-$ center near a
semiconductor-dielectric interface. We used material constants for the semiconductor side
corresponding to $\varepsilon_S=11.9$ (Si) and for the oxide side $\varepsilon_O=3.4$ (SiO$_2$).

\begin{figure}[tb]
\begin{center}
\includegraphics*[width=12cm]{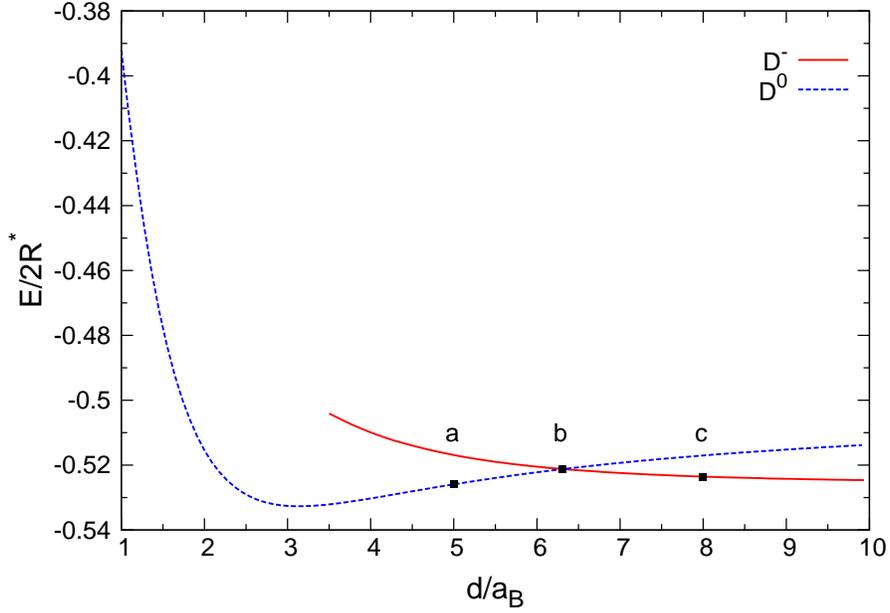}
\end{center}
\caption{(Color online) The ground state energy (solid red curve) of a $D^-$ center near a
semiconductor-dielectric interface vs. the donor-interface distance $d$. The energy for a neutral
donor ($D^0$) (dashed blue curve) near a semiconductor-dielectric interface is shown for
comparison. The black squares are the d-values for which we show the electron density in
Fig.~\ref{fig_eDen_diel}.} \label{en_diel}
\end{figure}

\begin{figure}[tb]
\begin{center}
\includegraphics*[width=12cm]{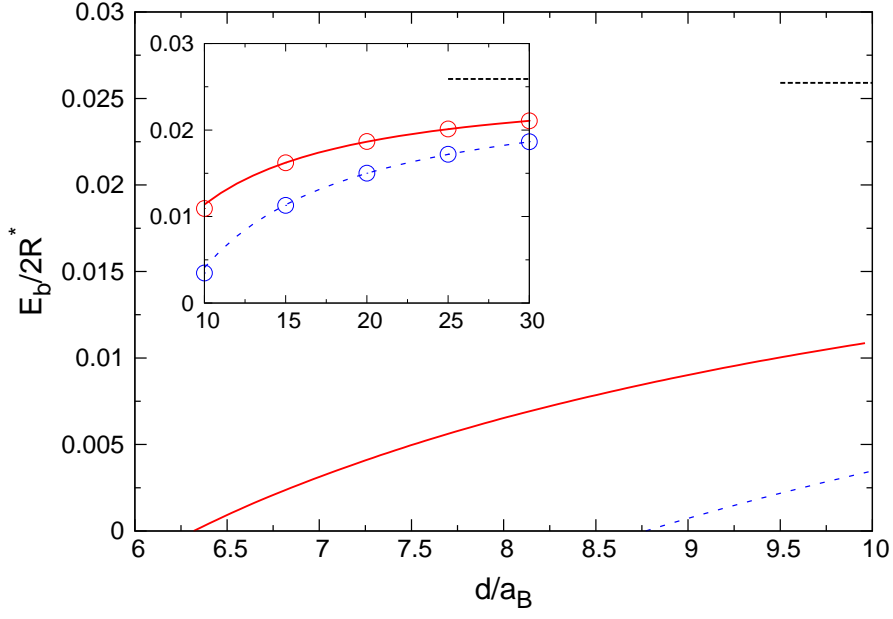}
\end{center}
\caption{(Color online) The binding energy for a $D^-$ center near a semiconductor-dielectric
(semiconductor-vacuum) interface vs. the donor distance $d/a_B$ indicated by solid red (dashed
blue) curve. The horizontal dashed black line shows the bulk limit of the binding energy. The
inset shows the behavior of the binding energy (symbols) for large $d$. The solid red (dashed
blue) curve is a fitting curve based on the equation $E/2R^*=0.0259-x a_B/d$ with the fitting
parameter $x=0.145$ ($x=0.218$).} \label{fig_BindingE_diel_vac}
\end{figure}

\begin{figure}[tb]
\begin{center}
\includegraphics*[width=13cm]{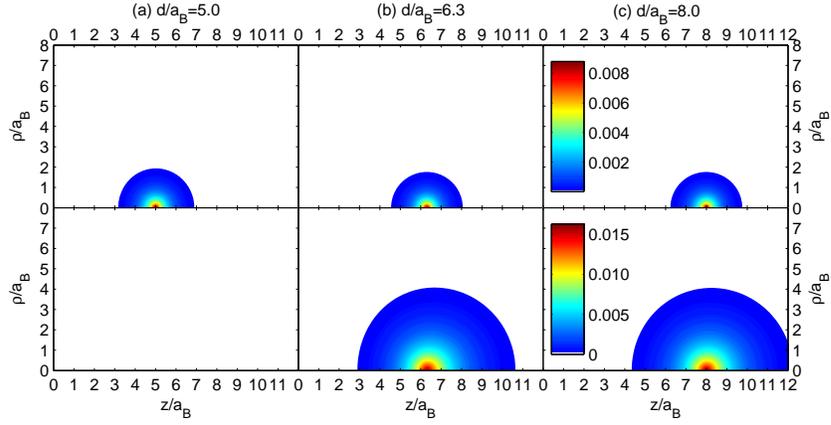}
\end{center}
\caption{(Color online) The electron density of the $D^-$ center near a semiconductor-dielectric
interface at donor distances $d/a_B=5$, 6.3 and 8 (see the black squares in Fig.~\ref{en_diel}).
The top three figures correspond to the inner electron density (when the outer electron is fixed
to its average position), whereas the bottom three figures are the densities of the outer electron
with the inner electron fixed to its average. Blue (red) area represents low (high) probabilities
and the white area represents almost zero probability.}
\label{fig_eDen_diel}\end{figure}

The potential energy between the particles is still given by Eq.~(\ref{eq_H}) with corresponding
value of $Q=0.556$. The results are presented in Fig.~\ref{en_diel} for the ground state energy
and in Fig.~\ref{fig_BindingE_diel_vac} for the binding energy. The second electron is bound to
the $D^0$ for $d \geq 3.5 a_B$ which compares to $d=3.2a_B$ when we ignore the image charges. This
can be explained by the fact that for the $D^-$ system near a semiconductor-dielectric interface
the two electrons induces two negative image charges while they are only attracted by one positive
image charge coming from the donor. In Fig.~\ref{en_diel} we show also the energy of the $D^0$ by
the dashed blue curve which corresponds to the situation where the second electron is at infinity.
Notice that both curves cross at $d=6.3a_B$ and for smaller $d$-values the $D^-$ state has a
higher energy and consequently the second electron will be unbound. The binding energy of the
second electron to the $D^0$ is plotted in Fig~\ref{fig_BindingE_diel_vac} by the solid red curve.
This result approaches the bulk result slowly and from below. The inset of
Fig.~\ref{fig_BindingE_diel_vac} shows the large d-behavior which is fitted by the curve
$E_b/2R^*=0.0259-0.145a_B/d$. In Fig.~\ref{fig_eDen_diel} we show the electron density of the
inner and outer electron at $d/a_B=5,$ 6.3 and 8, corresponding to the black squares marked in
Fig.~\ref{en_diel}. Notice that the outer electron is much further extended in space than the
inner electron and is pushed away from the interface.
\newline

\noindent \textbf{C. Semiconductor-vacuum interface}

For the $D^-$ center near a semiconductor-vacuum interface, the material constant is chosen to be
$\varepsilon_S=11.9$ for the semiconductor side and $\varepsilon_O=1$ for the vacuum side, leading
to $Q=0.845$.

The qualitative behavior for the energy of the $D^-$ and $D^0$ as function of $d$ is similar to
the one shown if Fig.~\ref{en_diel}. Due to the larger $Q$, the role of the image charges becomes
more important and the outer electron becomes bound for $d=4a_B$. The crossing point between the
$D^-$ and $D^0$ curves is pushed to $d=8.8a_B$. The binding energy is shown in
Fig.~\ref{fig_BindingE_diel_vac} by the dashed blue curve. Notice the strongly reduced binding
energy as compared to the Si/SiO$_2$ interface and even more so when we compare it with the
semiconductor-metal interface case. The large d-behavior is shown in the inset which could be
fitted by the curve $E_b/2R^*=0.0259-0.218a_B/d$.

\section{Conclusion} \label{sec4}

We proposed a variational approach to investigate the energetics and the wave function extend of
the spin singlet ground state of the $D^-$ system that is located near a semiconductor-metal or a
semiconductor-dielectric (vacuum) interface. As a trial function we used a modified Chandrasekhar
type wave function, which differs with the Chandrasekar variational two-electron wavefunction in
the following way: 1) it satisfy the boundary condition at the interface and 2) it takes into
account all the Coulomb interactions with the image charges. This makes the wavefunction no longer
spherical symmetric. This variational approach gives the well-known $d\rightarrow\infty$ limit.

We obtained a nonmonotonic behavior of the $D^-$ binding energy as function of donor position near
a semiconductor-metal interface with a local maximum for a donor distance from the interface of
about $3.8a_B$. For smaller $d$-values the binding energy decreases and for $d<1.1a_B$ the $D^-$
becomes unbound which is mostly a consequence of the boundary condition of the two electron
wavefunction at the interface. At large $d$ the outer electron is attracted to the interface
because of the positive total image charge. For a semiconductor-dielectric (vacuum) interface the
$D^-$ binding energy is strongly reduced and is a uniform increasing function of $d$. When the
donor moves towards the interface, the energy of the $D^-$ reduces and the system becomes unbound
for $d<3.5a_B$ ($d<4a_B$) near a Si/SiO$_2$ (Si/vacuum) interface. While for a neutral $D^0$
center near a semiconductor-metal (semiconductor-dielectric) interface the contribution of the
image terms to the binding energy is approximately given by the expression $\Delta E=1/4d$
($-Q/4d$) for large $d$. In the case of the $D^-$ system, due to the presence of the two
electrons, a complete compensation of such terms takes place.

The present calculation was done within the effective mass approximation which is expected to be
valid for the considered length scale $d>a_B$. For a donor very close to the interface, i.e.
within three monolayers, the deformation of the lattice close to the interface, i.e. strain
relaxation near the donor, may invalidate the effective mass approximation. We also neglected the
penetration of the electrons in the metal (and the dielectric) which for the obtained binding
energies is expected to be a good approximation. In order to go beyond the present effective mass
approximation one can use approaches such as the tight-binding approximation or approaches based
on the density functional theory.

\begin{acknowledgments} This work was supported by the Belgian Science Policy (IAP) and the
Brazilian Science Foundation CNPq. One of us (AAV) was supported by a fellowship from the Belgian
Federal Science Policy Office (IAP).
\end{acknowledgments}

\begin{figure}[tb]
\begin{center}
\includegraphics*[width=12cm]{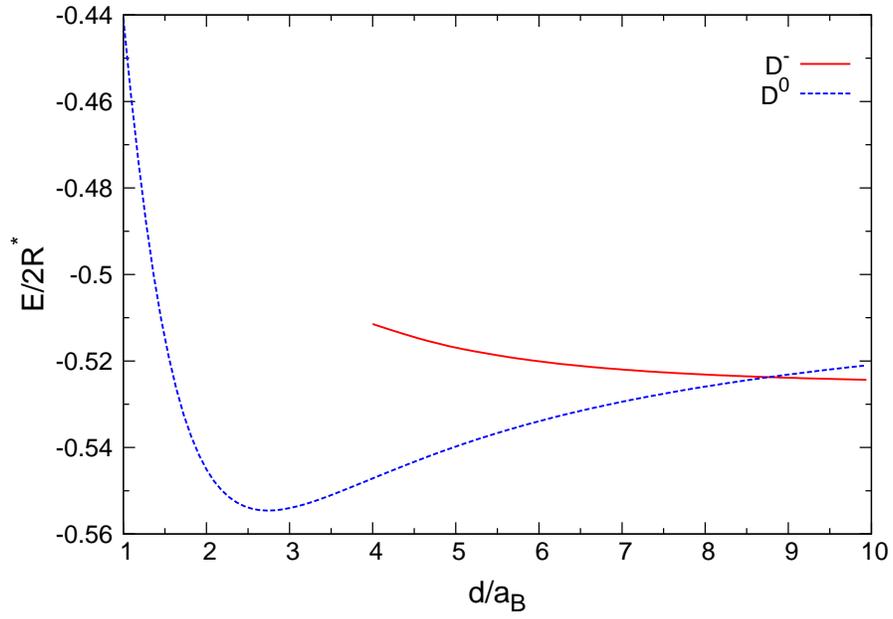}
\end{center}
\caption{(Color online) Similar as Fig.~\ref{en_diel} but now for the $D^-$ and $D^0$ near a
semiconductor-vacuum interface.} \label{fig_E0_vac}
\end{figure}

\end{document}